\documentclass[twocolumn,showpacs,preprintnumbers,amsmath,amssymb,prb,superscriptaddress]{revtex4}

\usepackage{graphicx}% Include figure files
\usepackage{dcolumn}% Align table columns on decimal point
\usepackage{bm}% bold math

\newcommand{\be}{\begin{equation}}
\newcommand{\ee}{\end{equation}}

\newcommand{\beq}{\begin{eqnarray}}
\newcommand{\eeq}{\end{eqnarray}}

\def\eq#1{(\ref{#1})}
\def\H1{\widehat{H}_1}

\newcommand{\pd}{\partial}

\begin{document}

\title{Persistent currents in multicomponent Tomonaga-Luttinger liquid: application to mesoscopic semiconductor ring with spin-orbit interaction }
   
\author{M. Pletyukhov}
\affiliation{Institut f\"ur Theoretische Festk\"orperphysik, Universit\"at 
Karlsruhe, D-76128 Karlsruhe, Germany}

\author{V. Gritsev}
\affiliation{D\'epartement de Physique, Universit\'e de Fribourg,
Chemin du Mus\'ee 3, CH-1700 Fribourg, Switzerland}

\begin{abstract}
We study persistent currents in semiconductor ballistic rings with spin-orbit 
Rashba interaction. We use as a working model the multicomponent Tomonaga-Luttinger liquid which arises due to the nonparabolic dispersion relations of electrons  in the  rings with rather strong spin-orbit coupling. This approach predicts some new characteristic features of persistent currents, which may be observed in experimental studies of semiconductor ballistic rings.   
\end{abstract}

%    73.23.Ra -  Persistent currents,
%    71.70.Ej -  Spin-orbit coupling,
%    73.23.Ad -  Ballistic transport
                          

\maketitle

\section{Introduction}

It has been known for a long time that due to the conservation of the electron phase coherence  an isolated metallic mesoscopic ring threaded by a magnetic flux may carry a persistent current\cite{BY}. This quantity, which equals to the derivative of free energy with respect to 
the flux, has been theoretically shown to exist in both ballistic rings\cite{Ku} and rings with disorder\cite{BIL}. The effect of impurities on persistent currents has been studied later on in a number of theoretical works\cite{CGRS,CRG,AE,OR,AGI,S}.

The experimentally measured currents\cite{Levy,Chandra,ex3} in diffusive metallic rings demonstrated a disagreement with the theoretically predicted values. This observation has initiated an intensive debate. Several authors have studied the effect of Coulomb interaction on persistent current\cite{AE,YF,BPM,MW,AB,CP,GS,Rozh,Zai}, either without or with disorder. There is still no total consensus on the simultaneous effect of impurities and electron-electron (e-e) interactions on persistent current. 

On the other hand, the experimental results for ballistic GaAs/GaAlAs-based rings\cite{MCB} were found to be in a good agreement with the theoretical predictions. 

For one-dimensional (1D) systems there is a variety of powerful methods which allow to treat e-e interactions. For instance, the persistent current in the Hubbard ring with a flux and Coulomb repulsion can be found from the Bethe Ansatz (BA) solution\cite{YF,Kus}. The effects of disorder in the Hubbard model have been studied in Ref.~\onlinecite{GS} by means of the bosonization technique. 

It is generally believed that the low-energy physics of 1D systems can be often  described using  the Tomonaga-Luttinger (TL) liquid concept\cite{H} (see Refs.~\onlinecite{voit,Gia} for the review). In particular, quantum wires made of semiconductor heterostructures have been found to demonstrate the TL features\cite{QW}. 
Therefore, the TL model appears to be a reasonable approximation for accounting for different effects in the 1D semiconductor heterostructures. 

In Ref.~\onlinecite{Loss} the persistent current has been studied within the TL model, and the particular emphasis has been made on the effects of electron number parity. The crucial significance of parity effects for the persistent current in 1D systems has been recognized quite a long time ago\cite{BY,CGRS}. In particular, it has been noticed\cite{Leg} that for a system of $N_e$ interacting spinless fermions in the TL phase the sign of a current, which reflects the properties of the ground state, is diamagnetic for odd $N_e$, while it is paramagnetic for even $N_e$.
 
For the case of spinful interacting fermions the nature of the ground state may change depending  both  on the strength of Coulomb repulsion and on the relative parity of spin-up and spin-down electron numbers. For instance\cite{YF}, in the Hubbard rings with an odd number of electrons  $N_e = N_{e \uparrow} + N_{e \downarrow}$ the current is paramagnetic around $\Phi =0$, and its period is half the flux quantum $\Phi_0 \equiv hc/e$. Meanwhile, in the Hubbard rings with even $N_e$ and low filling the current is diamagnetic near zero flux, whereas  at densities close to half-filling  the currents may become paramagnetic.

In the modern expanding field of spintronics the main research interest focuses basically on spin injection and spin detection in solid-state devices\cite{ZFS,Sil}. Semiconductor InAlAs/InGaAs-based heterostructures provide promising opportunities for spin manipulation, because in such materials spin-orbit (SO) splitting of  Rashba type\cite{Rashba,Sil} is rather large, and this allows for especially strong coupling of spin polarization and electric field. A naturally arising problem is to understand the effect of SO coupling on such a ground-state property as persistent current in a mesoscopic ring made of semiconductor material. Indeed, the issue of SO effects in noninteracting 1D ballistic rings has been already addressed in Refs.~\onlinecite{EGMO,gef,SGZ}. Surprisingly, the exact BA solution which takes into account the SO effects in the appropriately modified Hubbard model is also available\cite{SS}. Persistent currents in this system have been analyzed on the ground of the BA solution in Ref.~\onlinecite{FK}. The interplay of Coulomb repulsion and the SO coupling in combination with the parity effects leads to the remarkable features in the nature of current: although the SO interaction does not produce any effect in the case of a ring with the odd number $N_e$, the result for even-$N_e$ rings differs from that of  the standard Hubbard model without SO interaction. 

In the regime of low electron densities the 1D system of interacting electrons with SO coupling can also be described in terms of a {\it multicomponent} TL model. The persistent current in the SO-splitted  TL liguid has been studied in Ref.~\onlinecite{mosk} using the approach developed in Refs.~\onlinecite{Loss,FK}.

We have been discussing so far the effects of SO coupling which are caused by the relative {\it horizontal} shift of spin-up and spin-down {\it parabolic} dispersion curves. In such a situation the densities of states near the Fermi energy are the same in each branch, and therefore they can be characterized by a unique value of the Fermi velocity. However, it has been recently shown\cite{MB,egues,gover2} that the pre\-sen\-ce of strong SO interaction  can qualitatively change  the band structure of a 2D electron gas confined to the 1D geometry. This can be achieved by applying both a specific lateral confining potential in the plane of a 2D electron gas and the  Rashba potential perpendicular to this plane. The single particle electron
spectrum of thus effectively created ballistic quantum (quasi)-1D wire reveals quite a strong deviation from parabolic shape. The extent  of this deviation depends significantly on the relation between the strength of confining potential and the Rashba SO coupling constant. In order to capture the physics of the effective 1D wire with nonparabolic, spin-splitted spectrum it has been recently suggested to consider a modified TL model characterized by two different Fermi velocities\cite{MSB} $v_1 \neq v_2$. Without loss of generality, this model may be regarded as a multicomponent TL liquid introduced by Penc and S\'{o}lyom\cite{PS}.
 
The goal of this paper is to study the persistent currents in the class of the multicomponent TL liquids described above, emphasizing the features which can be potentially observed in experiment. In Sec.~\ref{disper} we review the derivation of the dispersion relations in the ring geometry performed in Refs.~\onlinecite{aronov,MMK,SGZ}, presenting it in an optimized form. In particular, we consider other profiles of the confining potential which allow to treat the radial part of the wavefuction without any further approximation. We also estimate the parameter $\eta = (v_1 - v_2)/(v_1 + v_2)$, which measures the asymmetry (nonparabolicity) of the single particle spectrum for the recently fabricated semiconductor ring with Rashba SO coupling\cite{klap2}. In Sec.~\ref{bosTL} we outline the basic steps of bosonization procedure and thus introduce our notations. We also quote the main results concerning the multicomponent TL liquids. We appreciate the importance of the zero-mode contributions to a persistent current and therefore take a special care of the zero-mode part of the bosonized TL Hamiltonian. In Secs.~\ref{perfiel} and \ref{perchem} we consider persistent currents in the cases of canonical and grand canonical ensembles, respectively. The important issue in these sections is the parity dependence of persistent currents, which in bosonization picture stems from the topological constraints (selection rules) imposed onto the topological excitations (zero modes). We observe the novel features of persistent currents caused by the nonzero value of $\eta$. The discussion of the results obtained is presented in Sec.~\ref{concl}.

\section{Mesoscopic rings with Rashba interaction: dispersion relations}
\label{disper}

The Hamiltonian for a quasi-1D ballistic ring with the Rashba SO coupling in polar coordinates\cite{aronov,MMK,SGZ} is a sum of the radial, angular and spin-orbit coupling terms
\beq
H_0^{rad} &=& -\frac{\hbar^2}{2 m} \left( \frac{\pd^2 }{\pd r^2} + 
\frac{1}{r} \frac{\pd}{\pd r}  \right) + V(r),  \\
H_0^{ang} &=& -\frac{\hbar^2}{2 m r^2} \left( -i \frac{\pd }{\pd \varphi} - q_{\Phi} \right)^2 , \\
H_{\mathrm{SO}} &=& \alpha_{\mathrm{R}} \frac{\sigma_r}{r}  \left( -i \frac{\partial }{\partial \varphi} - q_{\Phi} \right) + i\alpha_{\mathrm{R}}  \sigma_{\varphi} \frac{\pd }{\pd r}, 
\eeq
where $q_{\Phi} = \Phi/\Phi_0$, $m$ is an effective (band) mass of electron, $\alpha_{\mathrm{R}}$ is the Rashba coupling constant, $\sigma_r = \cos \varphi \sigma_1 +\sin \varphi \sigma_2$, $\sigma_{\varphi} = -\sin \varphi \sigma_1 +\cos \varphi \sigma_2$, and the Zeeman interaction is neglected. The  confining potential $V(r)$ can be modeled by a singular oscillator potential \cite{sing}
\be
V(r) = \frac{m \omega^2}{2} \left( r - \frac{a^2}{r}\right)^2, \label{so}
\ee
or by a hard-wall potential
\beq
V(r) &=& \frac{\hbar^2}{8 m r^2} \quad {\rm for}  \quad \left( a -\frac{d}{2} \right) < r < \left( a + \frac{d}{2} \right), \nonumber \\ V(r) &=& \infty \qquad \quad {\rm elsewhere}. \label{hw}
\eeq
In these expressions $a$  can be associated with a mean radius of the ring, and $l_{\omega} \equiv \sqrt{\hbar / m \omega}$ or $d$ -- with its mean width.  

We introduce dimensionless parameters $\lambda = a^2 / l_{\omega}^2$, $y = r/l_{\omega}$, $\alpha_0 = \alpha_{\mathrm{R}} m a/\hbar^2$, $h= H / \hbar \omega$, and rewrite the integrable $h_0^{rad}$ and nonintegrable parts $h_{\varphi,\sigma} \equiv h_0^{ang} + h_{\mathrm{SO}}$ in the new notations:
\beq
 h_0^{rad} &=& -\frac12 \left( \frac{\pd^2}{\pd y^2} + \frac{1}{y} \frac{\pd}{\pd y}\right) + v(y), \label{hrad} \\
 h_{\varphi,\sigma} &=& \frac{q^2}{2 y^2} + \frac{\alpha_0}{\sqrt{\lambda}} \left(\sigma_r \frac{q}{y} + i \sigma_{\varphi} \frac{\pd}{\pd y} \right) , \label{hang}
\eeq
where $q= -i \pd_{\varphi} - q_{\Phi}$. 

We can remove the dependence on $\varphi$ in \eq{hang} (which enters through  $\sigma_r$ and $\sigma_{\varphi}$) by a gauge transformation 
\be
h_{\varphi,\sigma} \to h'_{\varphi,\sigma} = e^{i \varphi \sigma_3 /2}h_{\varphi,\sigma} e^{-i \varphi \sigma_3 /2}, 
\label{gau1}
\ee
and therefore obtain
\beq
h'_{\varphi,\sigma}  &=& \frac{1}{2 y^2} \left( q - \frac12 \sigma_3 \right)^2
\nonumber \\
&+& \frac{\alpha_0}{\sqrt{\lambda}} \left(\sigma_1 \frac{q}{y} + i \sigma_2 \left(\frac{1}{2y} + \frac{\pd}{\pd y} \right)\right). \label{hang1}
\eeq
Note that due to the  transformation \eq{gau1} the periodic boundary conditions (at zero flux) for an angular wavefunction have changed to the anti-periodic ones.

The radial eigenfunctions  
\be
R_n (y) = \left[ \frac{\Gamma (n+1)}{\Gamma (n+ \lambda+ 1)} \right]^{1/2} 
\sqrt{2} y^{\lambda} e^{-y^2 /2} L_n^{\lambda} (y^2) , \label{sobas}
\ee
where $L_n^{\lambda}$ is the generalized Laguerre polynomial, and
\beq
R_n (y) &=& \sqrt{\frac{2}{y}} \sin \left[\pi (n+1) (y - \sqrt{\lambda} + 1/2) \right], \nonumber \\
& & \quad {\rm for} \quad  \sqrt{\lambda} -1/2 < y < \sqrt{\lambda} + 1/2, \nonumber \\
 R_n (y) &=& 0 \quad {\rm elsewhere}, \label{hwbas}
\eeq
 correspond to the confining potentials \eq{so} and \eq{hw}, respectively. In both cases they are labelled by $n=0,1,\ldots$ and normalized according to
\be
\int_0^{\infty} y R_m (y) R_n (y) dy = \delta_{mn}.
\ee
We would like to emphasize that in our consideration we do not make use of the approximation $\lambda \gg 1$, which is usually employed when the confining potential is modeled by a regular harmonic oscillator \cite{MMK,SGZ}. This becomes an important issue when the mean radius of the ring is comparable to its mean width.

We can find the spectra of \eq{hrad}
\be
\varepsilon_0^{rad} = 2n+1 \quad {\rm and} \quad \varepsilon_0^{rad} = \frac{\pi^2 (n+1)^2}{2}
\ee
for a singular oscillator and a hard-wall potentials, respectively, and the matrix elements $h'_{mn}$ of \eq{hang1} in the basis of $R_n (y)$:
\be
h'_{mn} = \frac{b_{mn}}{2} \left( q - \frac12 \sigma_3 \right)^2+ \frac{\alpha_0}{\sqrt{\lambda}} \left(\sigma_1 a_{mn} q + i \sigma_2 c_{mn} \right) . 
\label{hang2}
\ee 
Here we denote
\beq
a_{mn} &=& \int_{0}^{\infty} y R_m (y)
\left( \frac{1}{y} \right) R_n (y) dy, \label{amn} \\ 
b_{mn} &=& \int_{0}^{\infty} y R_m (y)
\left( \frac{1}{y^2} \right) R_n (y) dy, \label{bmn} \\
c_{mn} &=&\int_{0}^{\infty} y R_m (y)
\left( \frac{1}{2y} + \frac{\pd}{\pd y}\right) R_n (y) dy. \label{cmn}
\eeq
It is remarkable \cite{MMK} that the diagonal elements $c_{nn}$ identically vanish because they correspond to the momentum operator in polar coordinates.

Ideally, we have to diagonalize $h'_{mn}$ in the infinite basis $N \to \infty$ ($0 \leq m,n \leq N - 1 $). Practically, it can be only done approximately at some large but finite value of $N$. After having diagonalized $h'_{mn}$, we can perform another -- inverse of \eq{gau1} -- gauge transformation 
\be
h'_{diag} \to h''_{diag} = S h'_{diag} S^{-1}, \quad S = e^{-i \varphi \sigma_3 /2} \otimes 1_N 
\label{gau2}
\ee 
in order to restore the periodic boundary conditions for an angular wavefunction. The formal expression \eq{gau2} is equivalent to saying that the spin-up $\varepsilon'_{n \uparrow} (q)$ and spin-down $\varepsilon'_{n \downarrow} (q)$ dispersion curves in every band $n$ must be shifted in $q$ by $-1/2$ and $+ 1/2$, respectively.

To describe the main qualitative effects of spin-orbit coupling it is sufficient to consider the one-band ($N=1$) and two-band ($N=2$) approximations only. We calculate the matrix elements \eq{amn}-\eq{cmn} in the singular oscillator basis \eq{sobas}, as well as in the hard-wall potential basis \eq{hwbas}. In the former case we have
\beq
a_{00} &=& g (\lambda), \quad
a_{11} = \frac{\lambda +\frac34}{\lambda +1} g (\lambda), \label{a0s} \\
a_{01} &=& a_{10} = \frac{1}{2 \sqrt{\lambda + 1}} g (\lambda), \label{b0s1} \\
b_{00} &=&  b_{11} = \frac{1}{\lambda}, \quad b_{01} = b_{10} = 
\frac{1}{\lambda \sqrt{\lambda + 1}}, \label{b0s} \\ 
c_{01} &=& - c_{10} = -\frac{\lambda + \frac12}{\sqrt{\lambda +1}} g (\lambda),
\label{c0s}
\eeq
where $g (\lambda)$ is expressed through the $\Gamma$-function
\be
g (\lambda) = \frac{\Gamma (\lambda +\frac12)}{\Gamma (\lambda +1)}.
\ee
Respectively, in the latter case we have $c_{01} = - c_{10} = -8/3$ and
\beq
(a,b)_{nn} &=& \frac{\lambda^{-k/2}}{2 \pi} \!\! \int_{-\pi}^{\pi} \!\!
\frac{1 + (-1)^n \cos (n+1) y'}{\left( 1 + \frac{y'}{2 \pi \sqrt{\lambda}} \right)^k} dy', 
\label{offdiag1} \\
(a,b)_{01} &=& (a,b)_{10} \nonumber \\
&=& -\frac{\lambda^{-k/2}}{\pi} \int_{-\pi}^{\pi} 
\frac{\sin y' \cos \frac{y'}{2}}{\left( 1 + 
\frac{y'}{2 \pi \sqrt{\lambda}} \right)^k} dy', \label{offdiag2} 
\eeq
where $k=1,2$ correspond to $a$ and $b$, respectively. It is easy to compute the integrals in \eq{offdiag1} and  \eq{offdiag2} numerically. 

In the one-band approximation we find the dispersion relations
\be
h''_{00} = \varepsilon_{\pm} (q) =  \frac{b_{00}}{2} (q -q_{\mathrm{\Phi}}  
\mp q_{\mathrm{R}})^2 , 
\ee
where  $q_{\mathrm{R}} = \frac12 \left( \sqrt{1 + 4 \alpha_0^2 a_{00}^2 /\lambda b_{00}^2 } - 1 \right)$ is the Rashba angular momentum. The upper sign corresponds to the spin-up branch and the lower sign -- to the spin-down branch. In this approximation the effect of spin-orbit coupling shows up in the relative {\it horizontal} shift by $2 q_{\mathrm{R}}$  of the two {\it parabolic} dispersion curves.

\begin{figure}[t!]
\begin{center}
\includegraphics[width=6.1cm,angle=270]{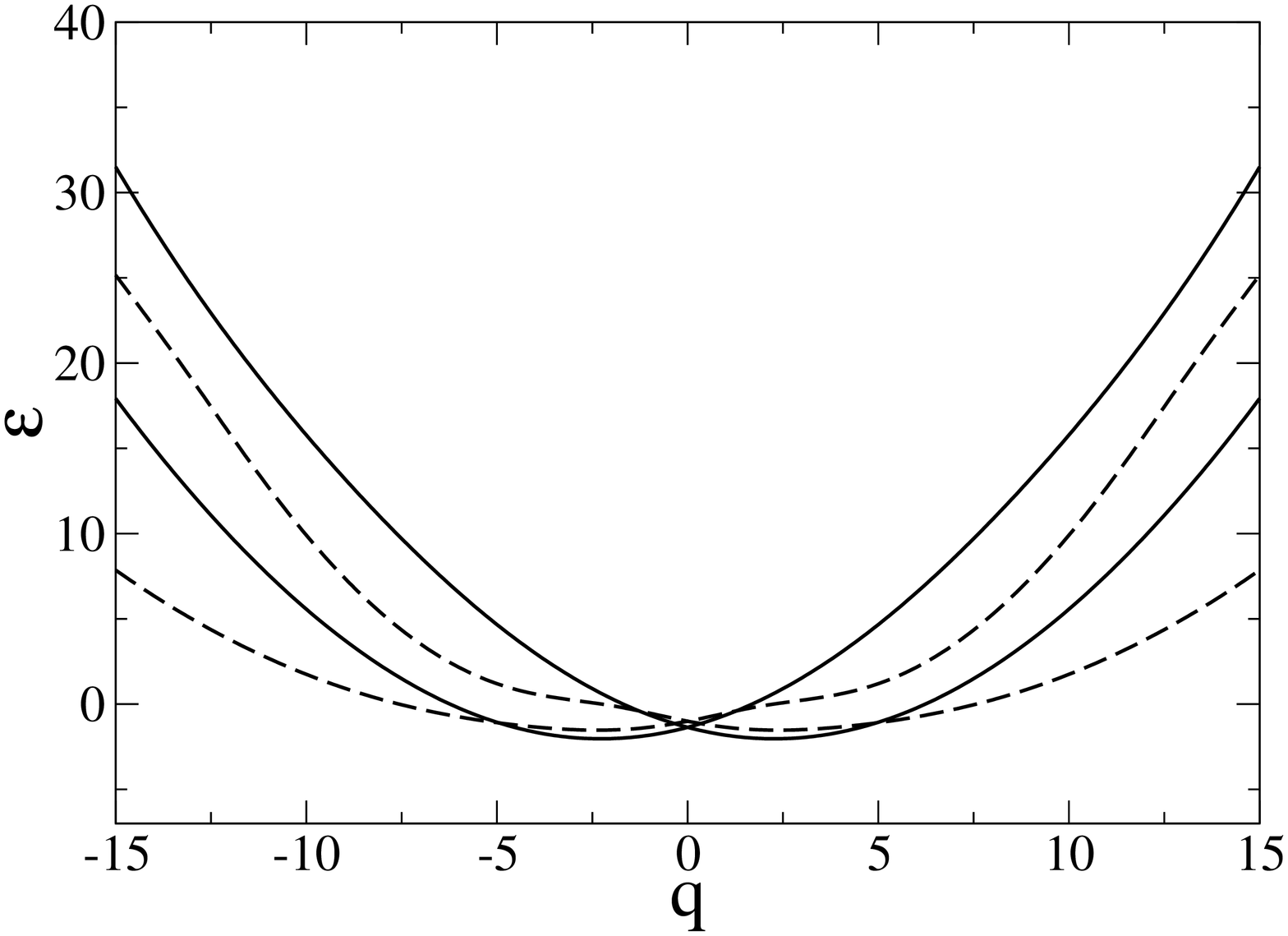}
\caption{The lowest radial band in the two-band approximation ($\alpha_0=3.0$). {\it Solid thick line:} hard-wall potential. {\it Dashed thick line:} singular oscillator potential. \label{sohwr3}}
\includegraphics[width=6.1cm,angle=270]{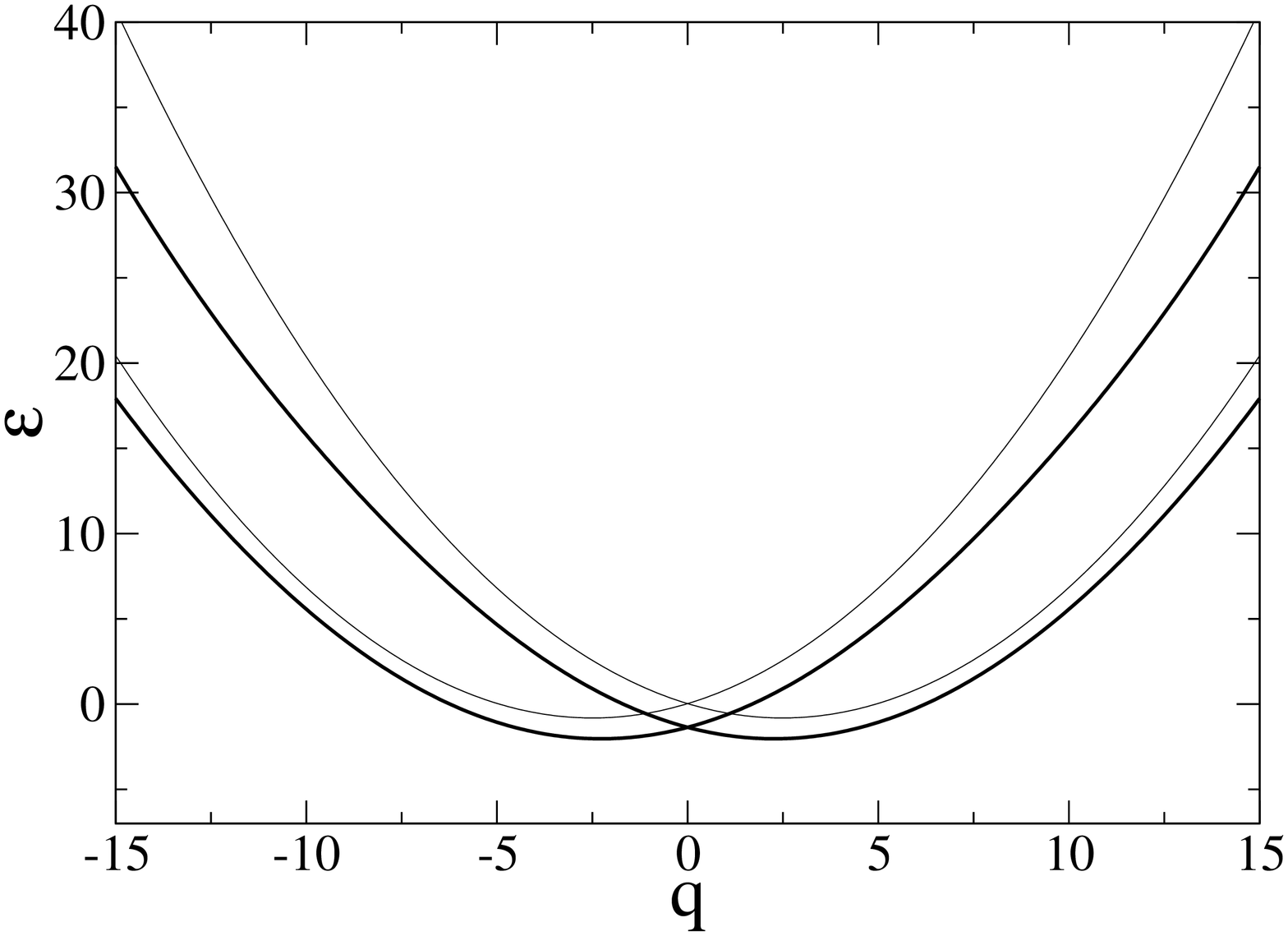}
\caption{The lowest radial band in the hard-wall potential ($\alpha_0=3.0$). {\it Solid thick line:} two-band approximation. {\it Solid thin line:} one-band approximation.\label{hwr3}}
\end{center}
\end{figure}

\begin{figure}[t!]
\begin{center}
\includegraphics[width=6.1cm,angle=270]{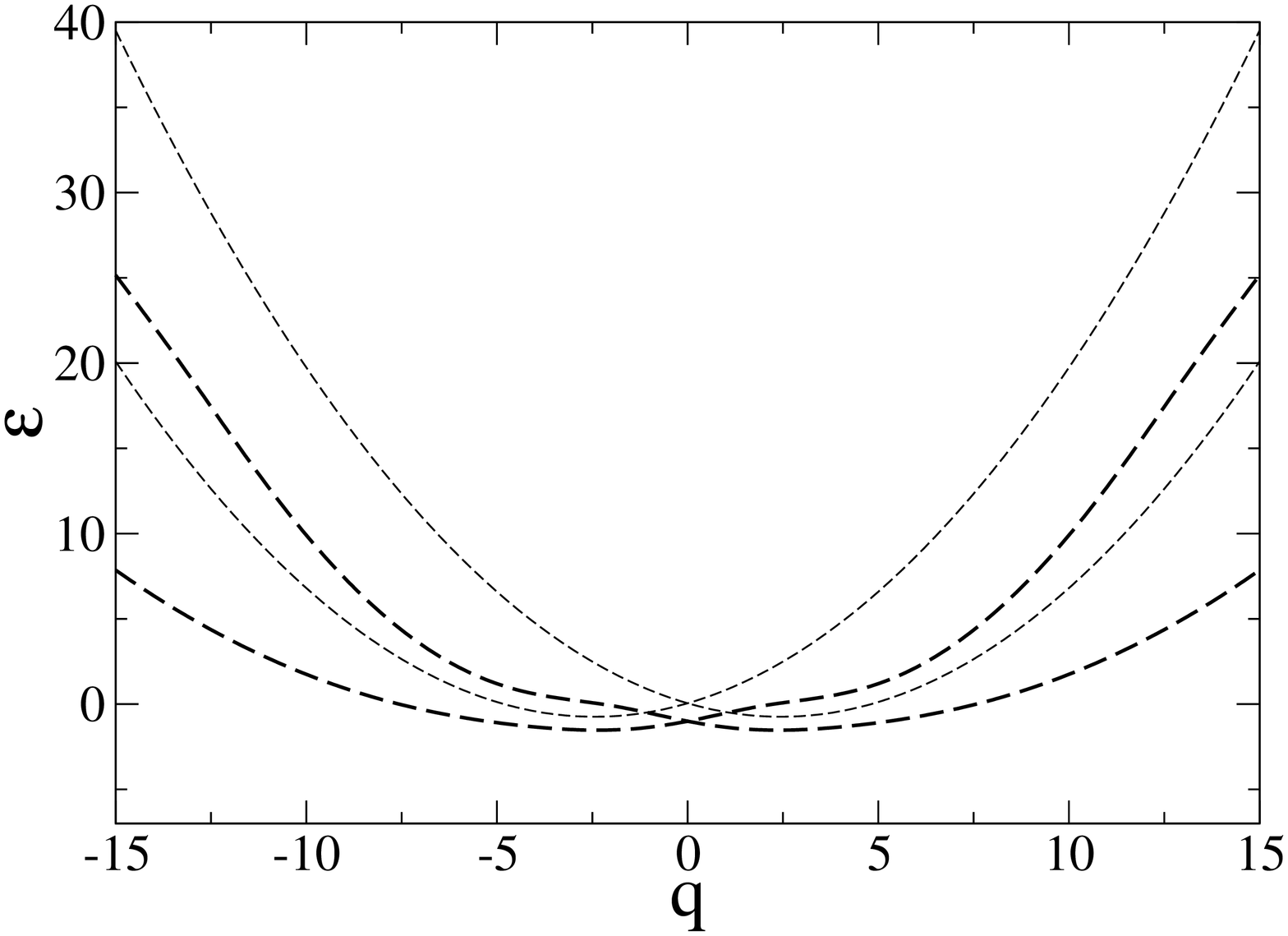}
\caption{The lowest radial band in the singular oscillator potential ($\alpha_0=3.0$). {\it Dashed thick line:} two-band approximation. {\it Dashed thin line:} one-band approximation.\label{sor3}}
\includegraphics[width=6.1cm,angle=270]{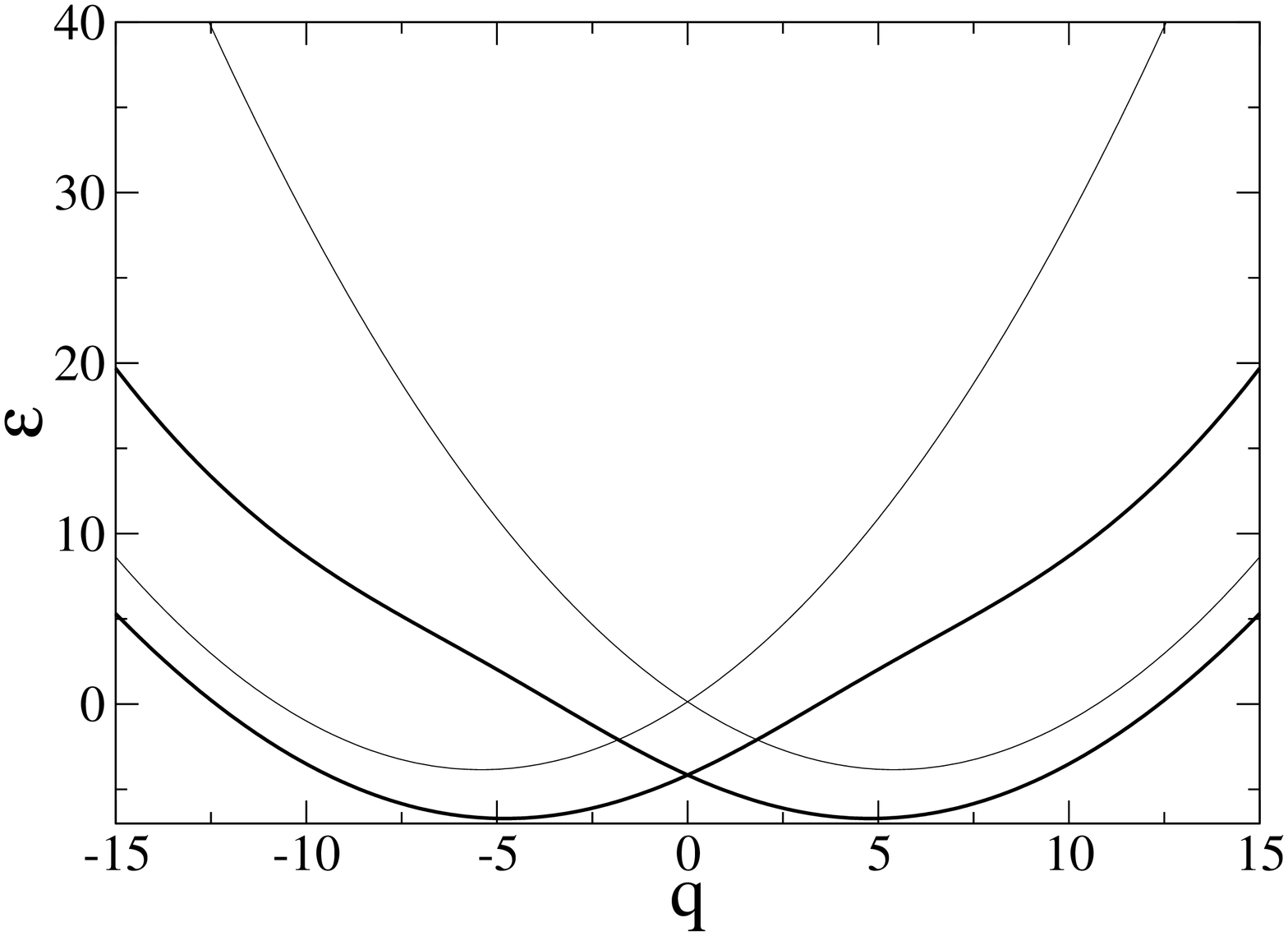}
\caption{The lowest radial band in the hard-wall potential ($\alpha_0=6.0$). {\it Solid thick line:} two-band approximation. {\it Solid thin line:} one-band approximation.\label{hwr6}}
\end{center}
\end{figure}

In the two-band approximation the analytical results are also available: we can find the solutions of the forth-order characteristic equation for $4 \times 4$-matrix \eq{hang2} and then shift them according to \eq{gau2}. However, it is easy to find the dispersion relations  in this case numerically. We use the following estimates for the parameters of the ballistic ring which has been studied in the recent experiment \cite{klap2}: $a=350$nm, $d=180$nm, mean free path $\simeq 1 \mu$m, $m \approx 0.042 m_e$.  The Rashba coupling constant $\alpha_{\mathrm{R}}$ for the heterostructure used in this experiment was $\approx 0.8 \cdot 10^{-11}$eVm. In other heterostructures it can reach even larger values (see, e.g. Ref.~\onlinecite{lev1}) in the range $\alpha_{\mathrm{R}} \approx 1 \div 6 \cdot 10^{-11}$eVm. With these parameters we can find $\lambda = 3.78$ and $\alpha_0$ in the range $2 \div 12$, as well as the matrix elements \\[0.2cm]
$a_{00} = 0.50 (0.52)$, $a_{11} = 0.47 (0.52)$, $a_{01} = 0.11 (0.05)$, \\[0.2cm]
$b_{00} = 0.26 (0.27)$, $b_{11} = 0.26 (0.28)$, $b_{01} = 0.12 (0.05)$, \\[0.2cm]
and $c_{01}=-0.97 (-2.67)$ in the singular oscillator (hard-wall) potential basis. The results of the band structure calculations are presented in Figs.~\ref{sohwr3}-\ref{hwr6}. In each figure the level $\varepsilon_0^{rad} (n=0)$ is set to zero. We plot the lowest radial band SO-splitted into the two subbands (spin-up and spin-down). In Fig.~\ref{sohwr3} we compare the results of the two-band approximation in the hard-wall and the singular oscillator potentials for the value of the Rashba constant $\alpha_0 =3.0$. In Fig.~\ref{hwr3} the results of the two-band and the one-band approximations in the hard-wall potential are presented at $\alpha_0 = 3.0$. In Fig.~\ref{sor3} the results of the two-band and the one-band approximations in the singular oscillator potential are presented at $\alpha_0 = 3.0$. In Fig.~\ref{hwr6} we compare the results of the two-band and the one-band approximations in the hard-wall potential at $\alpha_0 = 6.0$. So, we can see that the dispersion relations calculated in the ring geometry appear to be similar to the electron spectra in the {\it wire} geometry\cite{MSB,egues,gover2}. The common qualitative feature of the wire and the ring dispersion curves beyond the one-band approximation is their bending, i.e the deviation from the parabolic shape. It emerges already in the two-band approximation, and including higher bands into consideration would not qualitatively change the picture. The immediate consequence of the spectrum bending is the difference in Fermi velocities
\be
v_{\mathrm{F} \uparrow}^{right} \neq - v_{\mathrm{F} \uparrow}^{left}, \qquad v_{\mathrm{F} \downarrow}^{right} \neq - v_{\mathrm{F} \downarrow}^{left} .
\ee
However, due to the time-reversal symmetry expressed in the form $\varepsilon_{\uparrow} (q) = \varepsilon_{\downarrow} (-q)$ (at zero flux) we still have 
\be
v_{\mathrm{F} \uparrow}^{right} = - v_{\mathrm{F} \downarrow}^{left}, \qquad v_{\mathrm{F} \downarrow}^{right} = - v_{\mathrm{F} \uparrow}^{left} . 
\ee
Therefore, we have two distinct (in the absolute value) Fermi velocities $v_1 = v_{\mathrm{F} \uparrow}^{right}$ and $v_2 = v_{\mathrm{F} \downarrow}^{right}$. We introduce the notations $v_0 = (v_1 + v_2 )/2$, $\delta v = (v_1 - v_2)/2$ and $\eta = \delta v / v_0$. For $\alpha_0 = 6.0$ we make an estimate $\eta \sim 0.27$ in the hard-wall confining potential at the Fermi energy $\sim 1.48$ measured from $\varepsilon_0^{rad} (n=0)$.

Recently, there have been studied the effects of non-zero $\eta$ on two-point\cite{MSB} and four-point\cite{iucci} correlation functions of the TL liquid. In the next sections we would like to consider how the spectrum bending discussed above modifies the persistent currents.

\section{Bosonization of the multicomponent Tomonaga-Luttinger liquid}
\label{bosTL}

In this section we present the basic steps of bosonization procedure and the main issues of the TL liquid theory. In what follows we will use the notations 
\be
k = \frac{2 \pi q}{L} \quad  {\rm and} \quad 
x = \frac{L \varphi}{2 \pi} \nonumber
\ee 
instead of $q$ and $\varphi$ in order to employ the standard bosonization formulas and to have a transparent relation between the description of the ring and the wire geometries.

Linearization of the spectrum near the Fermi energy yields the second-quantized Hamiltonian
\beq
H &=& -i v_1 \int dx \left( \psi^{\dagger}_{R \uparrow} \pd_x \psi_{R \uparrow}  - \psi^{\dagger}_{L \downarrow} \pd_x \psi_{L \downarrow} \right) \nonumber \\
& & -i  v_2 \int dx \left( \psi^{\dagger}_{R \downarrow} \pd_x  \psi_{R \downarrow}  - \psi^{\dagger}_{L \uparrow} \pd_x \psi_{L \uparrow}\right) ,
\label{femh}
\eeq
where $\psi_{Rs} (x)$ and $\psi_{Ls} (x)$ are the right and the left components of the fermionic field 
\be
\psi_{s} (x) = \psi_{Rs} (x) + \psi_{Ls} (x), \quad s=\uparrow (+), \downarrow (-). 
\ee

The standard bosonization Ansatz reads\cite{schol}
 \be
\psi_{\eta s} (x) = \frac{F_{\eta s}}{\sqrt{L}} e^{i \eta \left( \frac{2 \pi}{L} N_{\eta s} - k_{\eta s} \right) x} \exp \left[ i \eta \sqrt{2 \pi} \phi_{\eta s} (x) \right] ,  \label{boson}
\ee
where $\eta = R(+), L(-)$, $N_{\eta s}$ are the topological excitations (zero modes) and $F_{\eta s}$ are the Klein factors. One can employ the mode expansion of the  bosonic fields 
\be
\phi_{\eta  s} (x) =  \frac{i}{\sqrt{L}} \sum_{k>0} \frac{1}{\sqrt{k}} e^{-\alpha k/2} \left[ e^{ i \eta k x} b_{k \eta s} - e^{- i \eta k x} b_{k \eta s}^{\dagger} \right] \nonumber
\ee
in terms of bosonic mode excitations
$b^{\dagger}_{k \eta s} (b_{k \eta s})$, $\alpha \to 0$ being a small cut-off parameter. We have also introduced the effects of a flux and a relative shift of the spin-up and spin-down dispersion curves by including the boundary term
\be
k_{\eta s}= \frac{k_1 + k_2}{2} + s \eta \frac{k_1 - k_2}{2} + \eta k_{\Phi} 
\ee
into the bosonization formula \eq{boson}. Here $k_1$ and $k_2$ are the linearization points for the right spin-up and the right spin-down branches, respectively. 

Bosonizing the Hamiltonian \eq{femh}, we obtain 
\be
H = \sum_{k>0} k \sum_{\eta s} | v_{\eta s} | b^{\dagger}_{k \eta s} b_{k \eta s} + \frac{\pi}{L} \sum_{\eta s} | v_{\eta s} | \widetilde{N}_{\eta s} (\widetilde{N}_{\eta s} +1),
\ee
where $| v_{R \uparrow } | = | v_{L \downarrow} | \equiv v_1$, $| v_{R \downarrow}| = | v_{L \uparrow } | \equiv v_2$, and
\be
\widetilde{N}_{\eta s} = N_{\eta s} - \frac{L}{2 \pi} k_{\eta s}.
\ee

The next step is to take into account the standard $g_{4c(s)}$ and $g_{2c(s)}$ interactions (forward scattering) which specify the Tomonaga-Luttinger model \cite{voit}. In our case we have the two-component TL liquid, and its continuous ($k \neq 0$) part can be diagonalized by the canonical transformation \cite{MSB} $A = \left( A_{ij} \right)$
\be
(b_{k R \uparrow}, b^{\dagger}_{k L \uparrow}, b_{k R \downarrow}, b^{\dagger}_{k L \downarrow}  )^T = A (d_{k1+}, d^{\dagger}_{k2+}, d_{k1-}, d^{\dagger}_{k2-} )^T , \nonumber
\ee 
where $A_{ij} = A_{ij} (g_{4c(s)}, g_{2c(s)} ; \delta v)$. Thus, we obtain 
\beq
H &=& \sum_{k >0} k \sum_{\nu = \pm} v_{\nu} \left( d^{\dagger}_{k 1 \nu} d_{k 1 \nu} + d^{\dagger}_{k 2 \nu} d_{k 2 \nu} \right)  \nonumber \\
&+& \frac{\pi}{4 L} \left[ v_c K_c \widetilde{J}_c^2 + v_s K_s \widetilde{J}_s^2 + \frac{v_c}{K_c} \widetilde{N}_c^2 +\frac{v_s}{K_s} \widetilde{N}_s^2 \right.
\nonumber \\
& & + \left. 2 \delta v \left( \widetilde{J}_c  \widetilde{N}_s  + \widetilde{J}_s  \widetilde{N}_c \right) + 4 v_0  \widetilde{N}_c + 4 \delta v  \widetilde{J}_s 
\right], \label{fembd}
\eeq 
where $v_{\pm}$ is expressed \cite{iucci} through the charge (spin) velo\-city $v_{c(s)}$, charge (spin) stiffness $K_{c(s)}$ and $\delta v$.

We note that the zero-mode part in \eq{fembd} (the second and the third lines) cannot be fully diagonalized by the 
{\it continuous} canonical transformation $A$, since the charge and spin current excitations
\beq
J_c &=& N_{R \uparrow} - N_{L \uparrow} +N_{R \downarrow} -N_{L \downarrow}, 
 \\
J_s &=& N_{R \uparrow} - N_{L \uparrow} -N_{R \downarrow} +N_{L \downarrow},
\eeq
as well as the charge  and spin number excitations
\beq
N_c &=& N_{R \uparrow} + N_{L \uparrow}+ N_{R \downarrow} +N_{L \downarrow}, \\
N_s &=& N_{R \uparrow} + N_{L \uparrow}- N_{R \downarrow} -N_{L \downarrow} 
\eeq
take integer values, and therefore they can be only transformed by a {\it modular} transformation which maps a grid onto itself. For this reason the grand partition function $\Xi_0$ corresponding to zero modes cannot be expressed in terms of the standard Jacobian theta functions \cite{grad} as it occurs in the spinless case \cite{Loss} and even in the spinful case with spin-orbit coupling \cite{mosk}. It is exactly the spectrum asymmetry parameter $\delta v$ that violates the factorization of $\Xi_0$ into different $J_c , J_s , N_c , N_s$ sectors. In the canonical ensemble the factorization of the partition function $Z_0$ is still possible, but the terms proportional to $\delta v$ would also cause some corrections.

In \eq{fembd} we have also introduced
\beq
\widetilde{J}_c &=& J_c - 4 q_{\Phi}, \quad \widetilde{J}_s = J_s - 4 \delta q ,\label{bound0}  \\
\widetilde{N}_c &=& N_c  - 4 q_0 , \quad \widetilde{N}_s = N_s, \label{bound}
\eeq
with 
\be
q_0 = \frac{L (k_1 + k_2 )}{4 \pi} \sim q_{\mathrm{F}}, \quad \delta q = \frac{L (k_1 - k_2 )}{4 \pi} \sim q_{\mathrm{R}}. \label{newq}
\ee
The similarity sign turns into the equality sign for the one-band approximation, i.e. when we deal with the parabolic dispersion relations. The first expression in \eq{bound0} also shows that the entire dependence on the magnetic flux is contained in the zero-mode part of the Hamiltonian \eq{fembd}, and therefore it is sufficient to calculate $Z_0 (\Xi_0 )$ in order to find the persistent currents in the  (grand) canonical ensemble. 

\section{Persistent currents: fixed number of electrons}
\label{perfiel}

In case of fixed number of electrons persistent current equals
\be
I (\Phi) = - \frac{d}{d \Phi} \Delta F (\Phi),
\ee
where $\Delta F (\Phi)$ is an oscillating part of the free energy.

In the following we adopt the units in which $\Phi_0 =1$ and introduce the characteristic scale for a current $I_0 \equiv 4 \pi  v_0 /L$. 

In the canonical ensemble the charge and spin number excitations are forbidden: $N_c = N_s \equiv 0$. Therefore, different topological sectors disentangle, and we find
\be
\Delta F (\Phi ) =- T \,\, \mathrm{ln} \sum_{\{J_c , J_s \}} e^{- H_0 (\Phi)/T}
\label{freen}
\ee
summing over $\{ J_c , J_s \}$ sectors. We have introduced the Hamiltonian
\be
H_0 ( \Phi ) = \frac{\pi}{4 L} \left[ v_c K_c \left( J_c - 4 q_{\Phi} \right)^2 + v_s K_s \left( J_s  -  4 q_{\mathrm{R}}^{eff} \right)^2 \right] 
\ee
along with the effective Rashba angular momentum [cf. \eq{newq}]
\be
q_{\mathrm{R}}^{eff} = \delta q +  \frac{2 q_0 - 1}{2 v_s K_s } \delta v .
\label{raseff}
\ee

The curly brackets in $\{J_c , J_s \}$ mean that the certain topological constraints are imposed onto the pair of values $J_c$, $J_s$. These constraints, or selection rules, are different for different parities of the electron number. This gives rise to the parity dependence of persistent currents\cite{BY,CGRS}. Before discussing it in further details, we are already able to state that the whole effect of the spectrum asymmetry in the case of fixed number of electrons is the renormalization of the Rashba angular momentum \eq{raseff}.

The topological constraints in question have been derived in Ref.~\onlinecite{Loss} for spinless TL liquid, and the case of the spin-$\frac12$ TL liquid has been considered later on by different authors\cite{faz,kriv,mosk}. In what follows we will use the formulation of Ref.~\onlinecite{faz}.

\subsection{$N_e = 4 N_0 +2$ and $N_e = 4 N_0$}

When the number of electrons in the ring equals to $N_e = 4 N_0 +2$, we have the following possible combinations of  $\{J_c , J_s \}$:
\be
\{4 n_c , 4 n_s \}, \,\,  \{4 n_c +2 , 4 n_s +2 \}.
\ee
The summations over $n_c$ and $n_s$ are unconstrained and run from $-\infty$ to $\infty$. Using the properties of the Jacobian theta functions (Appendix \ref{theta}), we can find that the persistent current $I^{(1)} (q_{\Phi} )$  equals to
\be
\pi T \frac{\theta'_3 (\pi q_{\Phi}, \gamma_c ) \theta_3 ( \pi q_{\mathrm{R}}^{eff}, \gamma_s)+ \theta'_4 (\pi q_{\Phi}, \gamma_c) \theta_4 ( \pi q_{\mathrm{R}}^{eff}, \gamma_s) }{\theta_3 ( \pi q_{\Phi}, \gamma_c) \theta_3 (\pi q_{\mathrm{R}}^{eff}, \gamma_s ) + \theta_4 (\pi q_{\Phi}, \gamma_c) \theta_4 (\pi q_{\mathrm{R}}^{eff}, \gamma_s)}, 
\label{in4n2}
\ee
where 
\be
\gamma_{c(s)} = e^{-\pi L T/ v_{c(s)} K_{c(s)} }. 
\ee

In the case $\delta v = 0$ we find the  agreement between \eq{in4n2} and the respective expression of Ref.~\onlinecite{mosk}. Considering further the noninteracting limit
\be
\gamma_c = \gamma_s = \gamma \equiv e^{-\pi L T/ v_0 },
\ee
we can simplify \eq{in4n2} down to 
\be
I^{(1)} (q_{\Phi}) = \sum_{n=1}^{\infty} \frac{4 \pi T (-1)^n}{\sinh (2 \pi^2 n T/ I_0 )} \cos ( 2 \pi n q_{\mathrm{R}}) \sin ( 2 \pi n q_{\mathrm{\Phi}}),
\ee
and recover the result of Ref.~\onlinecite{gef}.

For $N_e = 4 N_0$ we can find the persistent current by the mere shift of the argument 
\be
I^{(2)} (q_{\Phi} ) = I^{(1)} (q_{\Phi} + 1/2),
\ee
which is valid in both noninteracting\cite{gef} and interacting\cite{kriv} considerations.

\subsection{$N_e = 4 N_0 +1$ and $N_e = 4 N_0+3$}

When $N_e = 4 N_0 +1$ or $N_e = 4 N_0 +3$, i.e. for the odd electron number, we have to sum in \eq{freen} over the following combinations of $\{J_c , J_s \}$:
\beq
& & \{4 n_c +1 , 4 n_s +1 \}, \,\, \{4 n_c +1 , 4 n_s +3 \}, \nonumber \\
& & \{4 n_c + 3 , 4 n_s +1\}, \,\, \{4 n_c +3 , 4 n_s +3 \} . \label{topod}
\eeq
We can find then
\be
I^{(3)} ( q_{\Phi} ) = \sum_{n=1}^{\infty} \frac{4 \pi T}{\sinh (\pi n L T/ v_c K_c )} \sin ( 4 \pi n q_{\mathrm{\Phi}}).  \label{in4n3}
\ee
It is remarkable that the result does not depend on $q_{\mathrm{R}}^{eff}$, and therefore the spin-orbit coupling does not show up in this case at all. In the noninteracting limit $v_c = v_0$ and $K_c = 1$, and we are again in agreement with Ref.~\onlinecite{gef}.

From this perspective the formula (13) of  Ref.~\onlinecite{mosk} seems to be  incorrect. The probable source of error might be in the following: even though all possible topological configurations of the ground state  were correctly established (equivalently to \eq{topod}), not all of them were included in the computation of the free energy.  Therefore, the statement of  Ref.~\onlinecite{mosk} that for finite temperatures there might occur the spontaneous currents (at zero flux) is misleading. In contrast, according to \eq{in4n3} we always have $I^{(3)} (0) =0$.

\section{Persistent currents: fixed chemical potential}
\label{perchem}

Let us now consider the case when the mesoscopic ring is weakly coupled to the reservoir with chemical potential $\mu$. In this situation the charge and spin number excitations are allowed. For this reason the terms mixing $J$ and $N$ sectors in \eq{fembd} do not vanish, and therefore the consequences of the spectrum  asymmetry ($ \delta v \neq 0$) may become more diverse.

In order to find the persistent current
\be
I (\Phi , \mu) = -\frac{d}{d \Phi} \Delta \Omega (\Phi, \mu),
\ee
we need to calculate the oscillating part of the thermodynamic potential
\be
\Delta \Omega (\Phi , \mu ) = -T \,\, \mathrm{ln} \sum_{ \{J_c , N_c , J_s , N_s \}} e^{-  H_0 (\Phi , \mu)/T} ,
\ee
where the curly brackets in $\{J_c , N_c , J_s , N_s \}$ again mean the topological constraints, and 
\beq
&&  H_0 ( \Phi , \mu ) = \frac{\pi}{4 L} \! \left[ v_c K_c \widetilde{J}_c^2 + v_s K_s \widetilde{J}_s^2 + \frac{v_c}{K_c} \widetilde{N}_c^2 + \frac{v_s}{K_s} \widetilde{N}_s^2 \right.
\nonumber \\
&&+\left. 2 \delta v \left( \widetilde{J}_c  \widetilde{N}_s  + \widetilde{J}_s  \widetilde{N}_c \right) + 4 v_0  \widetilde{N}_c + 4 \delta v  \widetilde{J}_s \right] \! - \mu N_c. \label{hfimu}
\eeq
Due to $\delta v \neq 0$ different topological $J$ and $N$ sectors remain entangled. We can rewrite $16 H_0 ( \Phi , \mu )/I_0$ (up to some irrelevant additive constant term) in the form
\beq
& & \!\!  \lambda_c (J_c - z_{\Phi})^2 + 2 \eta (J_c - z_{\Phi}) (N_s - z_{\mathrm{B}}) + \nu_s (N_s - z_{\mathrm{B}})^2  \nonumber \\
&+& \!\!   \lambda_s (J_s - z_{\mathrm{R}} )^2 + 2 \eta (J_s - z_{\mathrm{R}} ) (N_c - z_{\mu}) + \nu_c (N_c - z_{\mu})^2   \nonumber 
\eeq
with $\eta$ and $I_0$ introduced in Secs.~\ref{disper} and \ref{perfiel}, respectively; 
\be
\lambda_{c(s)} = \frac{v_{c(s)} K_{c(s)}}{v_0}, \quad \nu_{c(s)} = \frac{v_{c(s)}}{K_{c(s)} v_0},
\ee
and
\beq
z_{\Phi} &=& 4 q_{\Phi}, \quad z_{\mathrm{R}} = 4 \left( \delta q - \frac{\eta (f_{\mu} + \nu_c )}{2 ( \nu_c \lambda_s - \eta^2 )}\right) , \\
z_{\mathrm{B}} &=& 0, \quad z_{\mu} =  4 \left( q_0 + \frac{f_{\mu} \lambda_s +  \eta^2}{2 ( \nu_c \lambda_s  - \eta^2 )}\right) .
\eeq
In the above expressions $f_{\mu} = \frac{\mu \pi L}{v_0} -1$ measures the difference between the chemical potential and the Fermi energy, and $z_{\mathrm{B}}$ is some argument, which would have been non-zero, had we included the Zeeman interaction.

We remark that the charging energy $E_c$ and the gate voltage $V_g$ are not explicitly included in \eq{hfimu}. We infer that they can be effectively taken into account by redefining $\lambda_c \to \lambda_c + 16 E_c /I_0$ and $\mu \to \mu + e V_g$, respectively.

In Appendix \ref{topcon} we introduce auxiliary functions $G (z_{\Phi} , z_{\mathrm{B}} , z_{\mathrm{R}} , z_{\mu})$ and $G' (z_{\Phi} , z_{\mathrm{B}} ,  z_{\mathrm{R}} , z_{\mu})$, which allow to conveniently express $\Xi (\Phi , \mu) = e^{-\Delta \Omega (\Phi , \mu )/T}$ and $I (\Phi , \mu )$ for different electron numbers $N_e$ in the ground state (at $T=0$ and $\Phi =0$). In the noninteracting limit and/or in the limit $\eta = 0$ the functions $G$ and $G'$ can be simplified and rewritten in terms of the standard Jacobian theta functions (due to the relations quoted in Appendix \ref{theta}). In general, they are expressed in terms of the Siegel theta functions\cite{sieg} (i.e. theta functions of higher degree), and therefore it is not straightforward to establish their asymptotics. However, for small enough temperatures the functions $G$ and $G'$ can be easily computed numerically.

\subsection{$N_e = 4 N_0 +2$ and $N_e = 4 N_0$ in the ground state}

The selection rules formulated in Ref.~\onlinecite{faz} are also applicable in our consideration. Thus, for $N_e = 4 N_0 +2$ electrons in the ground state we have to choose those $J$'s and $N$'s which satisfy the following requirements:
\begin{itemize}
\item[1)] $J_c$ and $N_c$ are either {\it simultaneously} even or  {\it simultaneously} odd. The same holds for $J_s$ and $N_s$;
\item[2)] $J_c \pm J_s + N_c \pm N_s$ takes values $\ldots, -4,0,4, \ldots$ 
\end{itemize}
In fact, this concisely formulated prescription assumes the summation over 16 different combinations of $\{J_c , N_c , J_s , N_s \}$ (see Appendix \ref{topcon}). Having properly performed it, we obtain
\beq
\Xi_1 (\Phi , \mu ) &=& G (z_{\Phi} , z_{\mathrm{B}} , z_{\mathrm{R}} , z_{\mu}), \label{grpf} \\
\Xi'_1 (\Phi , \mu ) &=& G' (z_{\Phi} , z_{\mathrm{B}} , z_{\mathrm{R}} , z_{\mu}), 
\eeq
and
\be
I^{(1)} (\Phi , \mu ) = 2 \pi T \frac{\Xi'_1 (\Phi, \mu )}{\Xi_1 (\Phi, \mu )}.
\label{curmu}
\ee

In the noninteracting limit $\lambda_{c(s)} = \nu_{c(s)} =1$ the  persistent current \eq{curmu} equals to
\be
4 \pi T \sum_{n=1}^{\infty} (-1)^n \left[ \frac{\sin \frac{\pi n}{2} z'_{\Phi} \cos \frac{\pi n}{2} z'_{\mathrm{R}}}{\sinh \frac{4 T \pi^2 n}{I_0 (1 +\eta)}} \! + \! \frac{\sin \frac{\pi n}{2} z'_{\mathrm{B}} \cos \frac{\pi n}{2} z'_{\mu}}{\sinh \frac{4 T \pi^2 n}{I_0 (1 -\eta)}}\right] \! ,
\label{nonper}
\ee
where 
\beq
z'_{\Phi} &=& z_{\Phi} + z_{\mathrm{B}}, \quad z'_{\mathrm{R}} = z_{\mathrm{R}} + z_{\mu} , \\
z'_{\mathrm{B}} &=& z_{\Phi} - z_{\mathrm{B}}, \quad z'_{\mu} = z_{\mathrm{R}} - z_{\mu} .
\eeq
Recalling that $z_{\mathrm{B}} =0$, we can easily establish the Fourier coefficients $I_n$ of the series
\be
\eq{nonper} = \sum_{n=1}^{\infty} I_n \sin 2 \pi n q_{\Phi}.
\ee
In the limit $T \to 0$ we find that $I_n$ equals to
\be
2 I_0 \frac{(-1)^n}{\pi n} \left[ \cos \frac{\pi n}{2} z_{\mathrm{R}} \cos \frac{\pi n}{2} z_{\mu}  + \eta \sin \frac{\pi n}{2} z_{\mathrm{R}} \sin \frac{\pi n}{2} z_{\mu}\right]. \nonumber
\ee
So, we obtain additional modulation due to $\eta \neq 0$.  

In the limit $\eta=0$ the grand partition function \eq{grpf} is proportional to
\be
\sum_{i=1}^4 \theta_i \left( \frac{\pi}{2} z_{\Phi}, \alpha_c^4 \right)
\theta_i \left( \frac{\pi}{2} z_{\mathrm{R}}, \alpha_s^4 \right)
\theta_i \left( \frac{\pi}{2} z_{\mu}, \beta_c^4 \right)
\theta_i \left( \frac{\pi}{2} z_{\mathrm{B}}, \beta_s^4 \right)
\label{etno}
\ee
with
\be
\alpha_{c(s)} = e^{-\frac{T \pi^2}{I_0 \lambda_{c(s)}}}, \quad \beta_{c(s)} = e^{-\frac{T \pi^2}{I_0 \nu_{c(s)}}}.
\label{abcs}
\ee
The coefficient of proportionality in \eq{etno} is omitted, since it does not depend on $z_{\Phi}$ and therefore becomes unimportant for the calculation of the persistent current \eq{curmu} in this limit. We note that the expression \eq{etno} coincides with that derived in  Ref.~\onlinecite{mosk}.

\begin{figure}
\begin{center}
\includegraphics[width=6.1cm,angle=270]{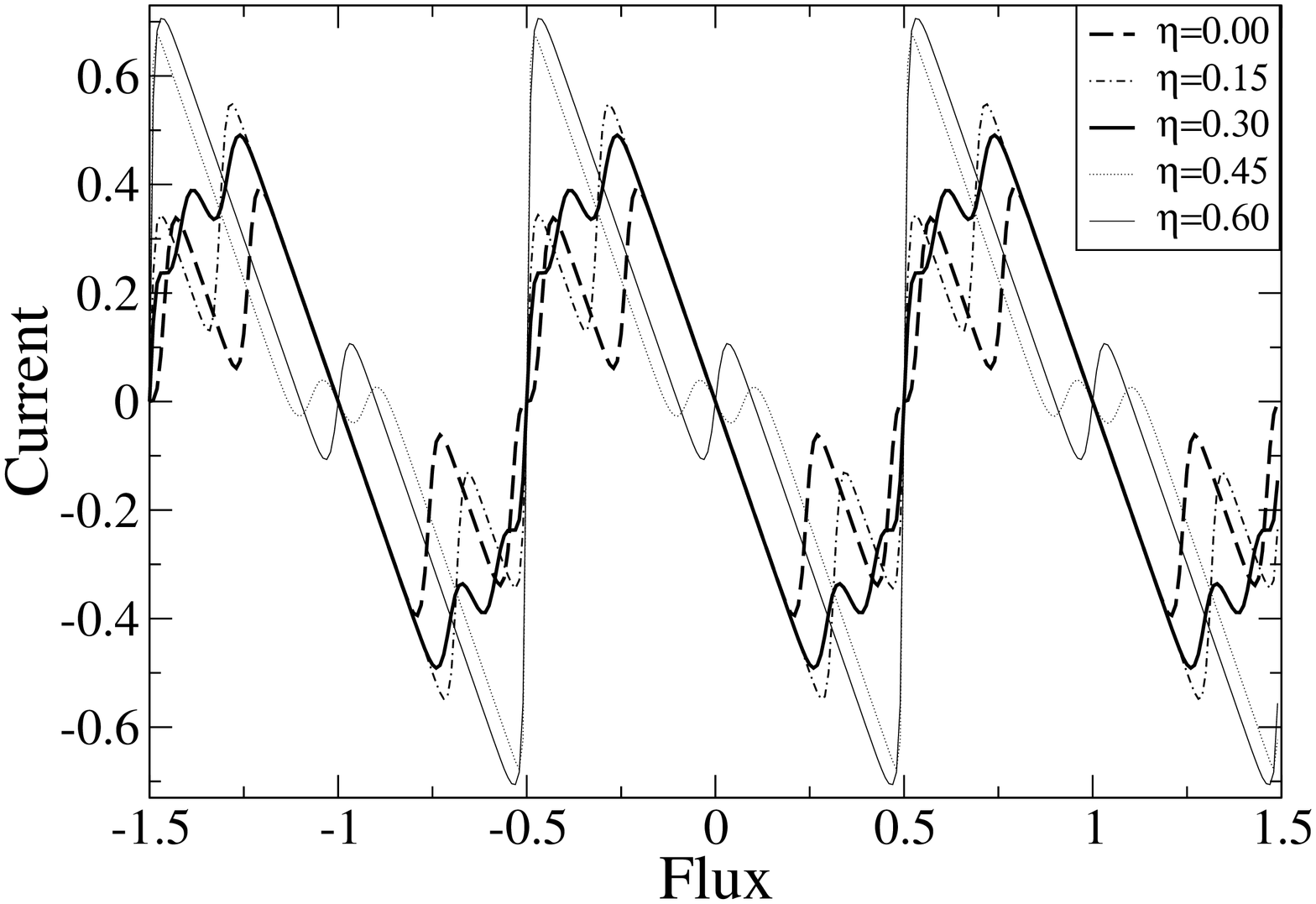}
\caption{Persistent currents $I/I_0$ vs. flux $\Phi/\Phi_0$ at $u = 0.4$ and different values of $\eta$ (fixed $\mu$, $N_e = 4 N_0 +2$). \label{evcur}} 
\includegraphics[width=6.1cm,angle=270]{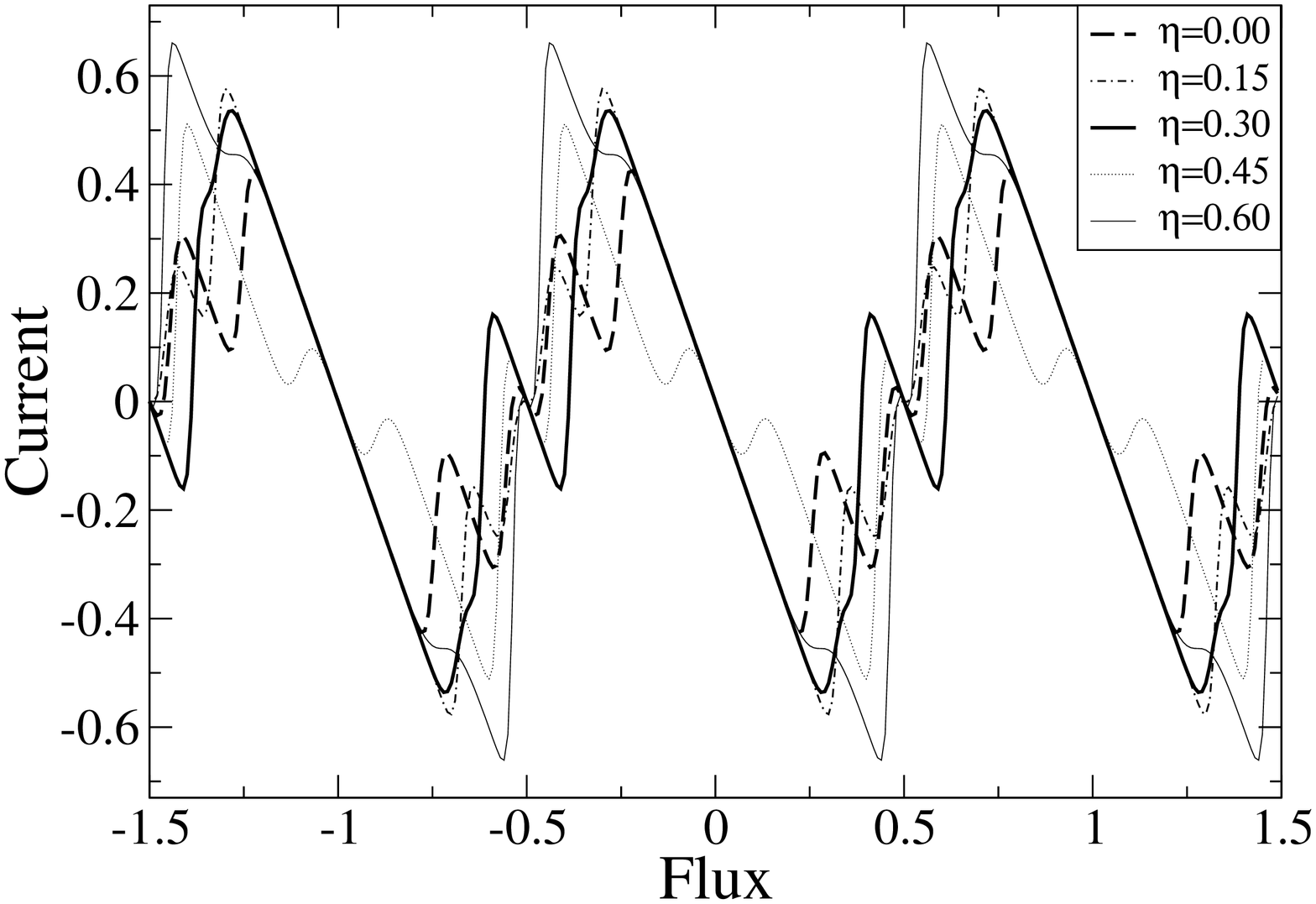}
\caption{Persistent currents $I/I_0$ vs. flux $\Phi/\Phi_0$ at $u = 0.0$ and different values of $\eta$ (fixed $\mu$, $N_e = 4 N_0 +2$).  \label{evcur1}}
\includegraphics[width=6.1cm,angle=270]{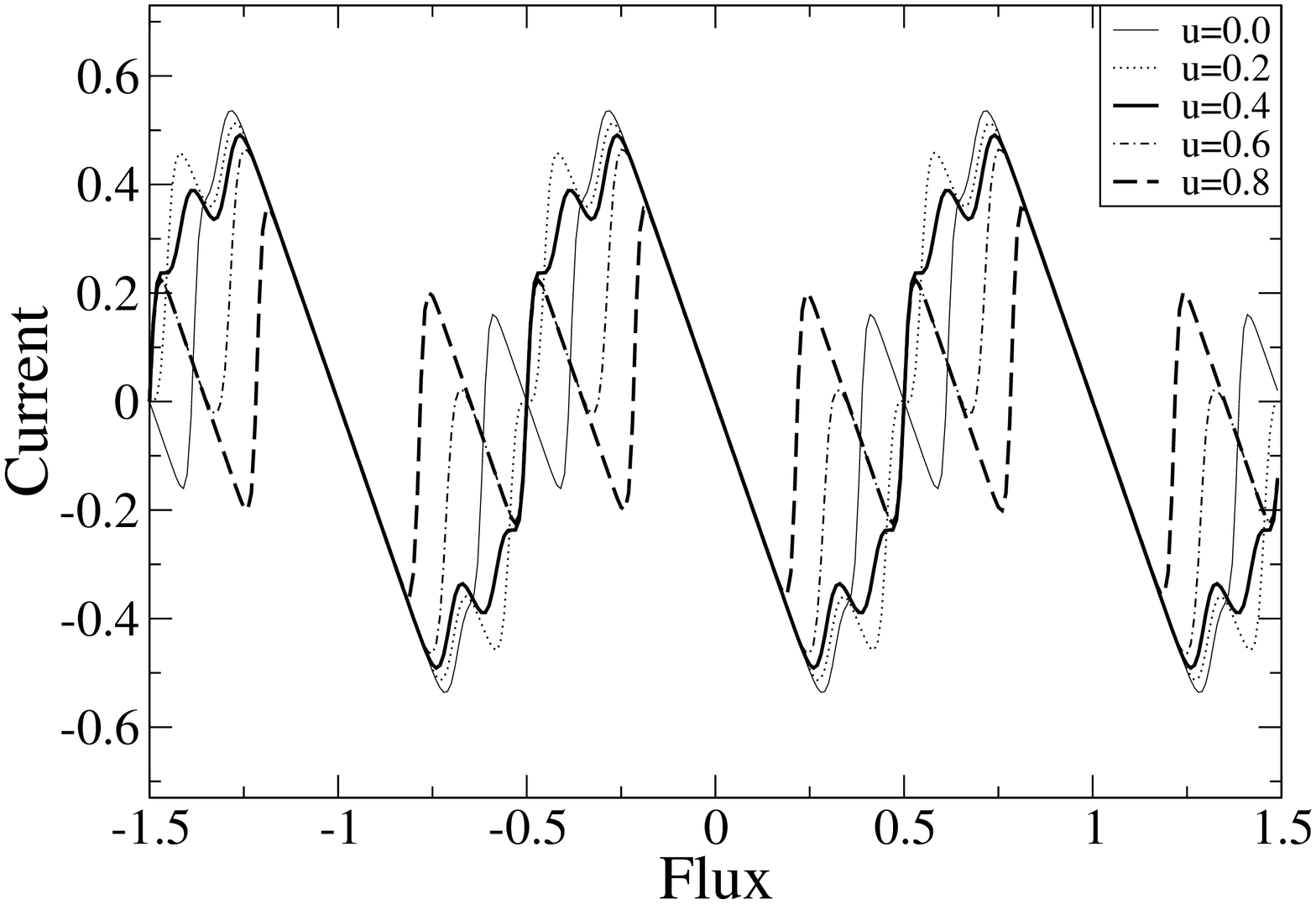}
\caption{Persistent currents $I/I_0$ vs. flux $\Phi/\Phi_0$ at $\eta=0.3$ and different values of $u$ (fixed $\mu$, $N_e = 4 N_0 +2$).  \label{evcur2}}
\end{center}
\end{figure}

In  Figs.~\ref{evcur}-\ref{evcur2} the persistent currents for the ring with $N_e = 4 N_0 +2$ electrons in the ground state (i.e. at $T=0$) are presented at $T=0.005 I_0$. To parameterize the e-e interactions we use the relation between the parameters of the TL model and the Hubbard model at low densities\cite{Gia}: $\lambda_c = \lambda_s = 1$, $\nu_c = 1+u$, $\nu_s = 1-u$, where $u= U/\pi v_0$ and $U$ is an interaction parameter in the Hubbard model. Other parameters are $\delta q = 0.15$, $q_0 = 0.04$, $f_{\mu} = 0.11$.

In Figs.~\ref{evcur} and \ref{evcur1} we show how the persistent current in  rings with  interacting ($u=0.4$) and noninteracting ($u=0.0$) electrons is modified when we vary the spectrum asymmetry parameter $\eta$. One can see that in the first case  increasing $\eta$ develops a paramagnetic ``tooth'' near $q_{\Phi} =0$, while in the second case altering $\eta$ changes the size of the diamagnetic ``tooth'' near $q_{\Phi} =1/2$. 

The dependence of the persistent current at $\eta=0.3$ on the interaction parameter $u$ is depicted in Fig.~\ref{evcur2}.

Of course, an alteration of other parameters $\delta q$, $q_0$, $f_{\mu}$ may also change the picture of the current, but it is obvious that the modification due to $\eta$ is remarkable.

The results for $N_e = 4 N_0$ are obtained by the shift in either $z_{\Phi}$ or $z_{\mu}$ (see, e.g. Ref.~\onlinecite{kriv}):
\beq
 \Xi_2 (\Phi , \mu ) &=& G (z_{\Phi} + 2, z_{\mathrm{B}} , z_{\mathrm{R}} , z_{\mu}) \nonumber \\
&\equiv& G (z_{\Phi} , z_{\mathrm{B}} , z_{\mathrm{R}} , z_{\mu}+2 ),
\eeq
and similarly for $I^{(2)} (\Phi , \mu )$.

\subsection{$N_e = 4 N_0 +1$ and $N_e = 4 N_0+3$ in the ground state}

First of all, we would like to remark that the ground state with the odd electron number ($N_e = 4 N_0 +1$ or $N_e = 4 N_0 +3$) can be justified only for repulsive interactions at very low temperatures: $T \ll U/L \sim I_0 u$ (see, e.g., Ref.~\onlinecite{kriv} for the discussion). In this case we can formulate the selection rules in terms of the function $G$ and thus obtain the following expression for the grand partition function:
\beq
\Xi_3 (\Phi , \mu ) &=& G (z_{\Phi} + 1, z_{\mathrm{B}} , z_{\mathrm{R}} +1, z_{\mu}) \nonumber \\
&+& G (z_{\Phi} +1, z_{\mathrm{B}} , z_{\mathrm{R}} - 1 , z_{\mu} ), 
\label{grpf3} \\
\Xi'_3 (\Phi , \mu ) &=& G' (z_{\Phi} + 1, z_{\mathrm{B}} , z_{\mathrm{R}} +1, z_{\mu}) \nonumber \\
&+& G' (z_{\Phi} +1, z_{\mathrm{B}} , z_{\mathrm{R}} - 1 , z_{\mu} ).
\eeq
Respectively, 
\be
I^{(3)} (\Phi , \mu ) = 2 \pi T \frac{\Xi'_3 (\Phi, \mu )}{\Xi_3 (\Phi, \mu )}.
\label{curmu3}
\ee

In the noninteracting limit the grand partition function \eq{grpf3} is proportional to
\beq
& & \theta_4\!\left(\!\frac{\pi}{2}z''_{\Phi},\!\gamma_{+}\!\right)
\!\theta_4\!\left(\!\frac{\pi}{2}z''_{\mathrm{B}},\!\gamma_{-}\!\right)
\!\theta_3\!\left(\!\frac{\pi}{2}z''_{\mathrm{R}},\!\gamma_{+}\!\right)
\!\theta_3\!\left(\!\frac{\pi}{2}z''_{\mu},\!\gamma_{-}\!\right) \nonumber \\
&+& \theta_3\!\left(\!\frac{\pi}{2}z''_{\Phi},\!\gamma_{+}\!\right)
\!\theta_3\!\left(\!\frac{\pi}{2}z''_{\mathrm{B}},\!\gamma_{-}\!\right)
\!\theta_4\!\left(\!\frac{\pi}{2}z''_{\mathrm{R}},\!\gamma_{+}\!\right)
\!\theta_4\!\left(\!\frac{\pi}{2}z''_{\mu},\!\gamma_{-}\!\right), \nonumber
\eeq
where $\gamma_{\pm} = e^{-\frac{4 T \pi^2}{I_0 (1 \pm \eta)}}$ and
\be
z''_{\Phi,\mathrm{R}} = \frac{z_{\Phi} + z_{\mathrm{R}}}{2} \pm \frac{z_{\mathrm{B}} + z_{\mu}}{2} , \,\,
z''_{\mathrm{B},\mu} = \frac{z_{\Phi} - z_{\mathrm{R}}}{2} \pm \frac{z_{\mathrm{B}} - z_{\mu}}{2} . \nonumber 
\ee

In the limit $\eta = 0$ the grand partition function \eq{grpf3} 
is proportional to
\beq
& & \theta_4\!\left(\!\frac{\pi}{2}z_{\Phi},\!\alpha_c^4\!\right)
\!\theta_3\!\left(\!\frac{\pi}{2}z_{\mathrm{B}},\!\beta_s^4\!\right)
\!\theta_4\!\left(\!\frac{\pi}{2}z_{\mathrm{R}},\!\alpha_s^4\!\right)
\!\theta_3\!\left(\!\frac{\pi}{2}z_{\mu},\!\beta_c^4\!\right) \nonumber \\
&+& \theta_3\!\left(\!\frac{\pi}{2}z_{\Phi},\!\alpha_c^4\!\right)
\!\theta_4\!\left(\!\frac{\pi}{2}z_{\mathrm{B}},\!\beta_s^4\!\right)
\!\theta_3\!\left(\!\frac{\pi}{2}z_{\mathrm{R}},\!\alpha_s^4\!\right)
\!\theta_4\!\left(\!\frac{\pi}{2}z_{\mu},\!\beta_c^4\!\right). \nonumber 
\eeq
If we put in this expression $z_{\mathrm{R}} = z_{\mathrm{B}} = 0$ (no spin-orbit coupling and no magnetic field) and $\alpha_s = \beta_s$, we obtain the formula equivalent to that derived in Ref.~\onlinecite{kriv}:
\beq
\Xi_3 (\Phi , \mu ) & \sim &  \left[ \theta_3  \left( \frac{\pi}{2} z_{\Phi}, \alpha_c^4 \right) \theta_4  \left( \frac{\pi}{2} z_{\mu}, \beta_c^4 \right) \right. \nonumber\\
& & \left. + \,\, \theta_4  \left( \frac{\pi}{2} z_{\Phi}, \alpha_c^4 \right) \theta_3  \left( \frac{\pi}{2} z_{\mu}, \beta_c^4 \right) \right] .
\eeq

\begin{figure}
\begin{center}
\includegraphics[width=6.1cm,angle=270]{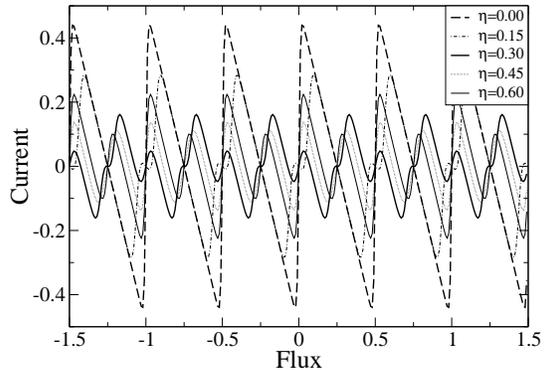}
\caption{Persistent currents $I/I_0$ vs. flux $\Phi/\Phi_0$ at $u=0.4$ and different values of $\eta$ (fixed $\mu$, odd $N_e$). \label{odcur}}
\end{center}
\end{figure}

In Fig.~\ref{odcur} we present the persistent currents at $u=0.4$ and different values of $\eta$, other parameters being the same as in the even-$N_e$ case. One can observe that in the presence of  $\eta \neq 0$ the persistent current is modified considerably. In particular, the effects of SO coupling show up in  the grand canonical ensemble in the ring with odd number of electrons in the ground state, in contrast to the ring with fixed odd number of electrons where such effects are absent [cf. \eq{in4n3}]. Due to $\eta \neq 0$ there also happen transitions between the diamagnetic and paramagnetic behavior at $q_{\Phi} = 0, \pm 1/4,\pm 1/2, \ldots$, and for some particular value of $\eta$ (which is close to $0.45$ for $u=0.4$) a period-halving occcurs (i.e. the period becomes equal to $q_{\Phi}/4$).

\section{Conclusions}
\label{concl}

We have demonstrated that the account of the higher radial subbands leads to the modification of the spectrum of electrons in the ring geometry. In particular, the dispersion relations obtained are characterized by two different Fermi velocities. Therefore, in order to describe the joint effect of the e-e interactions and spin-orbit coupling in ballistic rings it is reasonable to consider the multicomponent Tomonaga-Luttinger model. The interplay between the spin-orbit splitting, nonparabolicity of the spectrum ($\eta \neq 0$), e-e interactions and parity effects is reflected in the spectral properties of persistent currents. We have studied this within both canonical and grand canonical ensembles. Below we summarize the basic features of persistent currents caused by the subband nonparabolicity.

In Sec.~\ref{perfiel} we have shown that in the rings with the fixed electron number persistent currents for $\eta \neq 0$ do not change a lot compared to the case $\eta =0$. The reason for that is the  specific selection rule dictated by the particle conservation in the system. On the other hand, in the rings with fixed chemical potential we have observed considerable modifications. Thus, for the even number of electrons in the ground state we have obtained the generation of new harmonics and  complication of the current shape. We have observed how the picture changes if we vary the effective Coulomb interaction parameter $u$, while fixing the effective nonparabolicity parameter $\eta$, and vice versa, and found these changes remarkable. The modification of the current in the case of odd number of particles in the ground state appears to be even more drastic: varying $\eta$ we can perform the transition from diamagnetic to paramagnetic behavior at the flux values $\Phi = 0, \pm \Phi_0 /4, \pm \Phi_0 / 2, \ldots$ This makes the current essentially different from that derived from the parabolic dispersion relations for the same parity.

We therefore conclude that the nonparabolicity of the single particle electron spectrum produces a deep entanglement of the e-e repulsion and spin-orbit coupling parameters, and this leads to the new features of the current, which can be detected experimentally.

Recently, there has been the certain development in fabrication of toroidal carbon nanotubes\cite{carbon1}, and in the study of persistent currents in such systems\cite{carbon2}. Since the electronic spectrum in carbon nanotubes deviates considerably from the parabolic shape, we expect our results to be applicable in such realization as well, provided the selection rules are properly modified.   

%\vfill

\begin{acknowledgments}
We would like to thank Dionys Baeriswyl, George Japaridze and Oleg Zaitsev for fruitful communications. M.P. also acknowledges instructive discussions with Michele Governale and Sergei Sharov. M.P. was supported by the DFG Center for Functional Nanostructures at the University of Karlsruhe. V.G. was supported by the Swiss National Science Foundation.
\end{acknowledgments}

%\vfill

\appendix

\section{Properties of the Jacobian theta functions}
\label{theta}

We use the following definition of the Jacobian theta functions \cite{grad}
\beq
\theta_1 (z, \gamma) &=& 2 \gamma^{1/4} \! \sum_{n=0}^{\infty} (-1)^n \gamma^{n(n+1)} \sin (2n+1) z, 
\eeq
\beq
\theta_2 (z, \gamma) &=& 2 \gamma^{1/4} \sum_{n=0}^{\infty} \gamma^{n(n+1)} \sin (2n+1) z, 
\eeq
\beq
\theta_3 (z, \gamma) &=& 1+2\sum_{n=1}^{\infty} \gamma^{n^2} \cos 2  n z, 
\eeq
\beq
\theta_4 (z, \gamma) &=& 1+2\sum_{n=1}^{\infty} (-1)^n \gamma^{n^2} \cos 2 n z.
\eeq
Note that there exists the alternative definition which uses the different argument: $z \to \pi z$.

Functions $\theta_3$ and $\theta_4$ are periodic under the shift $z \to z + \pi$,  while $\theta_1$ and $\theta_2$ change their signs. Note also that
\be
\theta_3 \left( z + \frac{\pi}{2}, \gamma \right) = \theta_4 (z, \gamma ), \quad \theta_2 \left( z + \frac{\pi}{2}, \gamma \right) = -\theta_1 (z, \gamma ).
\nonumber
\ee

Making Poisson resummation, it is easy to prove the useful formulas
\beq
\sum_{k=-\infty}^{\infty} e^{-a (k+z)^2} 
&=& \sqrt{\frac{\pi}{a}} \theta_3 \left( \pi z , e^{-\pi^2 /a} \right), \\
\sum_{k=-\infty}^{\infty} (-1)^k e^{-a (k+z)^2} 
&=& \sqrt{\frac{\pi}{a}} \theta_2 \left( \pi z , e^{-\pi^2 /a} \right).
\eeq

From the relations
\be
\theta_{3,2} \left( 2 z , \gamma^4 \right) = \frac12 \left( \theta_3 (z , \gamma) \pm \theta_4 (z , \gamma) \right), 
\ee
it is easy to deduce that
\beq
& & \theta_3 \left( 2 z_1 , \gamma_1^4 \right) \theta_3 \left( 2 z_2 , \gamma_2^4 \right) + \theta_2 \left( 2 z_1 , \gamma_1^4 \right) \theta_2 \left( 2 z_2 , \gamma_2^4 \right ) \nonumber \\ 
&=& \frac12 \theta_3 (z_1 , \gamma_1) \theta_3 (z_2 , \gamma_2) + \frac12 \theta_4 (z_1 , \gamma_1) \theta_4 (z_2 , \gamma_2); \qquad \quad \label{t33} \\
& & \theta_3 \left( 2 z_1 , \gamma_1^4 \right) \theta_3 \left( 2 z_2 , \gamma_2^4 \right) - \theta_2 \left( 2 z_1 , \gamma_1^4 \right) \theta_2 \left( 2 z_2 , \gamma_2^4 \right) \nonumber \\
&=& \frac12 \theta_3 (z_1 , \gamma_1) \theta_4 (z_2 , \gamma_2) + \frac12 \theta_4 (z_1 , \gamma_1) \theta_3 (z_2 , \gamma_2). \qquad \quad \label{t44}
\eeq
Moreover, for $\gamma_1 = \gamma_2 \equiv \gamma$ 
\beq
\eq{t33} &=& \theta_3 \left( z_1 + z_2 , \gamma^2 \right) \theta_3 \left( z_1 - z_2 , \gamma^2 \right), \\
\eq{t44} &=& \theta_4 \left( z_1 + z_2 , \gamma^2 \right) \theta_4 \left( z_1 - z_2 , \gamma^2 \right).
\eeq
There also holds the following identity
\beq
 & & \sum_{i=1}^4 \theta_i (z_1 , \gamma) \theta_i (z_2 , \gamma) \theta_i (z_3 , \gamma) \theta_i (z_4 , \gamma) \nonumber \\
& & = 2 \theta_3 (z''_1 , \gamma) \theta_3 (z''_2 , \gamma) \theta_3 (z''_3 , \gamma) \theta_3 (z''_4 , \gamma),
\eeq
where
\be
z''_{1,2} = \frac{z_1 + z_2}{2} \pm \frac{z_3 + z_4}{2} , \,\,
z''_{3,4} = \frac{z_1 - z_2}{2} \pm \frac{z_3 - z_4}{2} .  \nonumber
\ee

In Ref.~\onlinecite{grad} one can also find  useful expressions for the logarithmic derivatives $\theta'_i (z,\gamma)/\theta_i (z, \gamma)$.

\section{Topological constraints for fixed chemical potential}
\label{topcon}

The topological constraints formulated in Ref.~\onlinecite{faz} for $N_e = 4 N_0 +2$ electrons in the ground state lead to the following possible combinations of $\{  J_c ,  N_c ,  J_s ,  N_s \}$:
\beq
& & \{ 4 n_c , 4 m_c , 4 n_s , 4 m_s \}, \nonumber \\ 
& & \{ 4 n_c +2 , 4 m_c +2 , 4 n_s +2 , 4 m_s +2 \}, \nonumber \\
& & \{ 4 n_c +2 , 4 m_c +2 , 4 n_s , 4 m_s \}, \nonumber \\
& & \{ 4 n_c , 4 m_c , 4 n_s +2 , 4 m_s +2 \}, \nonumber \\
& & \{ 4 n_c , 4 m_c +2 , 4 n_s , 4 m_s +2 \}, \nonumber \\
& & \{ 4 n_c , 4 m_c +2 , 4 n_s +2 , 4 m_s  \}, \nonumber \\
& & \{ 4 n_c +2 , 4 m_c  , 4 n_s , 4 m_s +2 \}, \nonumber \\
& & \{ 4 n_c +2 , 4 m_c , 4 n_s +2, 4 m_s \}, \nonumber \\
& & \{ 4 n_c + 1, 4 m_c + 1, 4 n_s + 1, 4 m_s + 1 \}, \nonumber \\  
& & \{ 4 n_c + 3, 4 m_c + 3, 4 n_s + 3, 4 m_s + 3 \}, \nonumber \\
& & \{ 4 n_c + 1, 4 m_c + 1, 4 n_s + 3, 4 m_s + 3 \}, \nonumber \\ 
& & \{ 4 n_c + 3, 4 m_c + 3, 4 n_s + 1, 4 m_s + 1 \}, \nonumber \\
& & \{ 4 n_c + 1, 4 m_c + 3, 4 n_s + 1, 4 m_s + 3 \}, \nonumber \\ 
& & \{ 4 n_c + 1, 4 m_c + 3, 4 n_s + 3, 4 m_s + 1 \}, \nonumber \\
& & \{ 4 n_c + 3, 4 m_c + 1, 4 n_s + 3, 4 m_s + 1 \}, \nonumber \\ 
& & \{ 4 n_c + 3, 4 m_c + 1, 4 n_s + 1, 4 m_s + 3 \}. 
\eeq
Therefore, we have to perform 16 different {\it unconstrained} summations over $n_c$, $m_c$, $n_s$, $m_s$ from $-\infty$ to $\infty$.

Let us define for convenience the functions [cf. Ref.~\onlinecite{sieg}]
\beq
f_1 (z_{\Phi},z_{\mathrm{B}}) &=& \sum_{n_c , m_s =-\infty}^{\infty} e^{- I_0 h_1 (z_{\Phi},z_{\mathrm{B}})/T}, \label{f1} \\
f_2 (z_{\mathrm{R}}, z_{\mu}) &=& \sum_{n_s , m_c= -\infty}^{\infty} e^{- I_0 h_2 (z_{\mathrm{R}}, z_{\mu})/T}, \label{f2}
\eeq
where
\beq
h_1 (z_{\Phi},z_{\mathrm{B}}) &=& \lambda_c \left( n_c - \frac{z_{\Phi}}{4} \right)^2 + \nu_s \left( m_s - \frac{z_{\mathrm{B}}}{4} \right )^2 \nonumber \\
& & + \,\, 2 \eta \left( n_c - \frac{z_{\Phi}}{4} \right) \left( m_s - \frac{z_{\mathrm{B}}}{4} \right ), \nonumber \\
h_2 (z_{\mathrm{R}}, z_{\mu}) &=& \lambda_s \left(n_s - \frac{z_{\mathrm{R}}}{4} \right)^2 + \nu_c \left( m_c - \frac{z_{\mu}}{4} \right)^2 \nonumber \\
& & + \,\, 2 \eta \left(n_s - \frac{z_{\mathrm{R}}}{4} \right) \left( m_c - \frac{z_{\mu}}{4} \right) , \nonumber
\eeq
and we must demand $\nu_s \lambda_c \! - \! \eta^2 \! > \! 0$ and $\nu_c \lambda_s \! - \! \eta^2 \! > \! 0$ to ensure the convergence of the series in \eq{f1} and \eq{f2}. Note that the functions $f_1$ and $f_2$ have period 4 in each argument.

If we introduce
\beq
g (z_{\Phi},z_{\mathrm{B}}, z_{\mathrm{R}}, z_{\mu}) &=& \left[ f_1 (z_{\Phi},z_{\mathrm{B}}) + f_1 (z_{\Phi}+2,z_{\mathrm{B}}+2) \right] \nonumber \\
& \times & \left[ f_2 (z_{\mathrm{R}}, z_{\mu}) + f_2 (z_{\mathrm{R}}+2, z_{\mu}+2)\right] \nonumber \\
&+& \left[ f_1 (z_{\Phi},z_{\mathrm{B}}+2) + f_1 ((z_{\Phi}+2,z_{\mathrm{B}}) \right] \nonumber \\
& \times & \left[ f_2 (z_{\mathrm{R}}, z_{\mu}+2) + f_2 (z_{\mathrm{R}}+2, z_{\mu})\right] , \nonumber 
\eeq
then it becomes easy to see that the grand partition function equals to
\beq
& & G (z_{\Phi},z_{\mathrm{B}}, z_{\mathrm{R}}, z_{\mu})  \equiv 
g (z_{\Phi},z_{\mathrm{B}}, z_{\mathrm{R}}, z_{\mu}) \nonumber \\ 
& & \quad + \,\, g (z_{\Phi}+1,z_{\mathrm{B}}+1, z_{\mathrm{R}}+1, z_{\mu}+1) .
\label{big}
\eeq

We can also define
\beq 
f'_1 (z_{\Phi},z_{\mathrm{B}}) &=& \frac{I_0}{2T} \sum_{n_c , m_s} \left[ \lambda_c \left( 4 n_c - z_{\Phi} \right) + \eta \left( 4 m_s - z_{\mathrm{B}} \right ) \right]  \nonumber \\
& & \qquad \qquad \times e^{- I_0 h_1 (z_{\Phi},z_{\mathrm{B}})/T},
\eeq
as well as $g'$ and $G'$ replacing $f_1$ by $f'_1$ in the above definitions.

In the noninteracting limit
\beq
h_1 (z_{\Phi},z_{\mathrm{B}}) &=& \frac{1+\eta}{2} \left( n_c + m_s - \frac{z_{\Phi} + z_{\mathrm{B}}}{4} \right)^2  \nonumber \\
&+&  \frac{1-\eta}{2} \left( n_c - m_s - \frac{z_{\Phi} - z_{\mathrm{B}}}{4} \right)^2. \nonumber
\eeq
So, we can transform $n_c$, $m_s$ by a modular transformation. Similarly, we can proceed with  $h_2$ and $n_s$, $m_c$.

In the limit $\eta =0$ the sum $f_1 (z_{\Phi},z_{\mathrm{B}}) + f_1 (z_{\Phi}+2,z_{\mathrm{B}}+2)$ is proportional to
$$
\theta_3 \left( \frac{\pi}{4} z_{\Phi}, \alpha_c \right)
\theta_3 \left( \frac{\pi}{4} z_{\mathrm{B}}, \beta_s \right)+
\theta_4 \left( \frac{\pi}{4} z_{\Phi}, \alpha_c \right)
\theta_4 \left( \frac{\pi}{4} z_{\mathrm{B}}, \beta_s \right),
$$
while $f_1 (z_{\Phi}+2,z_{\mathrm{B}}) + f_1 (z_{\Phi}+2,z_{\mathrm{B}})$ is proportional to
$$
\theta_3 \left( \frac{\pi}{4} z_{\Phi}, \alpha_c \right)
\theta_4 \left( \frac{\pi}{4} z_{\mathrm{B}}, \beta_s \right)+
\theta_3 \left( \frac{\pi}{4} z_{\Phi}, \alpha_c \right)
\theta_4 \left( \frac{\pi}{4} z_{\mathrm{B}}, \beta_s \right),
$$
where $\alpha_c$ and $\beta_s$ are given by \eq{abcs}. Similar relations hold for $f_2$. Using \eq{t33} and \eq{t44}, one can establish the expression \eq{etno}.

For the odd number of electrons in the ground state one can also establish the selection rules and calculate the grand partition function. In terms of the function $G$ it is presented in \eq{grpf3}. An alternative way to express $\Xi_3 (\Phi, \mu)$ is to introduce
\beq
& & \widetilde{G} (z_{\Phi},z_{\mathrm{B}}, z_{\mathrm{R}}, z_{\mu})  \equiv 
\tilde{g} (z_{\Phi},z_{\mathrm{B}}, z_{\mathrm{R}}, z_{\mu}) \nonumber \\ 
& & \quad + \,\, \tilde{g} (z_{\Phi}+1,z_{\mathrm{B}}+1, z_{\mathrm{R}}+1, z_{\mu}+1) ,
\label{big1}
\eeq
where
\beq
\tilde{g} (z_{\Phi},z_{\mathrm{B}}, z_{\mathrm{R}}, z_{\mu})  
&=&  g (z_{\Phi}+1,z_{\mathrm{B}}, z_{\mathrm{R}}+1, z_{\mu}) \nonumber \\
&+&  g (z_{\Phi}+1,z_{\mathrm{B}}, z_{\mathrm{R}}-1, z_{\mu}) \nonumber \\ 
&=& \tilde{f}_1 (z_{\Phi},z_{\mathrm{B}}) \tilde{f}_2 (z_{\mathrm{R}}, z_{\mu})
\label{small1}
\eeq
and
\beq
\tilde{f}_1 (z_{\Phi},z_{\mathrm{B}}) &=& f_1 (z_{\Phi}+1,z_{\mathrm{B}}) + f_1 (z_{\Phi}+1,z_{\mathrm{B}}+2) \nonumber \\
&+& f_1 (z_{\Phi}+3,z_{\mathrm{B}}) + f_1 (z_{\Phi}+3,z_{\mathrm{B}}+2), 
\nonumber \\
\tilde{f}_2 (z_{\mathrm{R}}, z_{\mu}) &=& f_2 (z_{\mathrm{R}}+1, z_{\mu}) + f_2 (z_{\mathrm{R}}+1, z_{\mu}+2) \nonumber \\
&+& f_2 (z_{\mathrm{R}}+3, z_{\mu}) + f_2 (z_{\mathrm{R}}+3, z_{\mu}+2).
\nonumber
\eeq

One can show that  the functions $\tilde{f}_1$ and $\tilde{f}_2$ have period 2 in each argument and equal to
\beq
\tilde{f}_1 (z_{\Phi},z_{\mathrm{B}}) &=& \sum_{\tilde{n}_c , \tilde{m}_s =-\infty}^{\infty} e^{- I_0 \tilde{h}_1 (z_{\Phi},z_{\mathrm{B}})/T}, \\
\tilde{f}_2 (z_{\mathrm{R}}, z_{\mu}) &=& \sum_{\tilde{n}_s , \tilde{m}_c= -\infty}^{\infty} e^{- I_0 \tilde{h}_2 (z_{\mathrm{R}}, z_{\mu})/T},
\eeq
where
\beq
\tilde{h}_1 (z_{\Phi},z_{\mathrm{B}}) &=& \frac14 \left[ \lambda_c \left( \tilde{n}_c - \frac{z_{\Phi}+1}{2} \right)^2 + \nu_s \left( \tilde{m}_s - \frac{z_{\mathrm{B}}}{2} \right )^2 \right. \nonumber \\
& & \quad \left. + \,\, 2 \eta \left( \tilde{n}_c - \frac{z_{\Phi}+1}{2} \right) \left( \tilde{m}_s - \frac{z_{\mathrm{B}}}{2} \right ) \right] , \nonumber \\
\eeq
\beq
\tilde{h}_2 (z_{\mathrm{R}}, z_{\mu}) &=& \frac14 \left[ \lambda_s \left( \tilde{n}_s - \frac{z_{\mathrm{R}}+1}{2} \right)^2 + \nu_c \left( \tilde{m}_c - \frac{z_{\mu}}{2} \right)^2 \right. \nonumber \\
& & \quad \left. + \,\, 2 \eta \left(\tilde{n}_s - \frac{z_{\mathrm{R}}+1}{2} \right) \left( \tilde{m}_c - \frac{z_{\mu}}{2} \right) \right]. \nonumber
\eeq

In the noninteracting limit
\beq
\tilde{h}_1 (z_{\Phi},z_{\mathrm{B}}) \!\! &=& \!\! \frac{1+\eta}{8} \left( \tilde{n}_c + \tilde{m}_s - \frac{z_{\Phi} + z_{\mathrm{B}}+1}{2} \right)^2  \nonumber \\
&+& \!\! \frac{1-\eta}{8} \left( \tilde{n}_c - \tilde{m}_s - \frac{z_{\Phi} - z_{\mathrm{B}}+1}{2} \right)^2 \!\! , \nonumber
\eeq
and we transform $\tilde{n}_c$, $\tilde{m}_s$ by a modular transformation. Similarly, we  proceed with  $\tilde{h}_2$ and $\tilde{n}_s$, $\tilde{m}_c$.

The limit $\eta =0$ for $\tilde{f}_1$ and $\tilde{f}_2$ is straightforward.

\vfill

%\newpage


\begin{thebibliography}{99}
\bibitem{Leg}
This statement is known as  Legett's theorem, see Ref.~\onlinecite{Loss} for the proof in the case of the TL liquid.
\end{thebibliography}
\end{document}